\documentclass[aps,prl,twocolumn,floats,letterpaper,footinbib,showpacs]{revtex4}

\usepackage{graphicx}
\usepackage[intlimits,centertags]{amsmath}
\usepackage{amssymb,amsfonts}
\usepackage[pdftex]{hyperref}

\hypersetup{pdfauthor=Markus Ahlers,pdftitle=Anomalous Anisotropies of Cosmic Rays from Turbulent Magnetic Fields,colorlinks=true,linkcolor=blue,citecolor=blue,filecolor=blue,urlcolor=blue}

\begin{document}

\title{Anomalous Anisotropies of Cosmic Rays from Turbulent Magnetic Fields}
\author{Markus Ahlers}
\affiliation{WIPAC \& Department of Physics, University of Wisconsin--Madison, Madison, Wisconsin 53706, USA}

\begin{abstract}
The propagation of cosmic rays (CRs) in turbulent interstellar magnetic fields is typically described as a spatial diffusion process. This formalism predicts only a small deviation from an isotropic CR distribution in the form of a dipole in the direction of the CR density gradient or relative background flow. We show that the existence of a global CR dipole moment necessarily generates a spectrum of higher multipole moments in the local CR distribution. These {\it anomalous} anisotropies are a direct consequence of Liouville's theorem in the presence of a local turbulent magnetic field. We show that the predictions of this model are in excellent agreement with the observed power spectrum of multi-TeV CRs.
\end{abstract}


\maketitle

{\it Introduction.}---The arrival directions of Galactic cosmic rays (CRs) are highly isotropic. This is expected from a diffusive propagation of CRs in the interstellar medium, where the effective scattering in turbulent magnetic fields randomizes the particle momenta over time. Diffusion theory (including also convective and dissipative processes) provides an excellent description of Galactic CR fluxes and their chemical abundances, {\it e.g.}~\citep{Blasi:2011fi}. In this framework the only deviation from an isotropic CR arrival direction is in the form of a weak dipole anisotropy. The phase and strength of this dipole is expected to be a combined effect of the relative motion of the solar system with respect to the frame where CRs are isotropic~\citep{Compton:1935} and the density gradient of CRs in the direction of their sources~\citep{Erlykin:2006ri,Blasi:2011fm,Pohl:2012xs}.

\begin{figure}[b]\centering
\includegraphics[width=\linewidth]{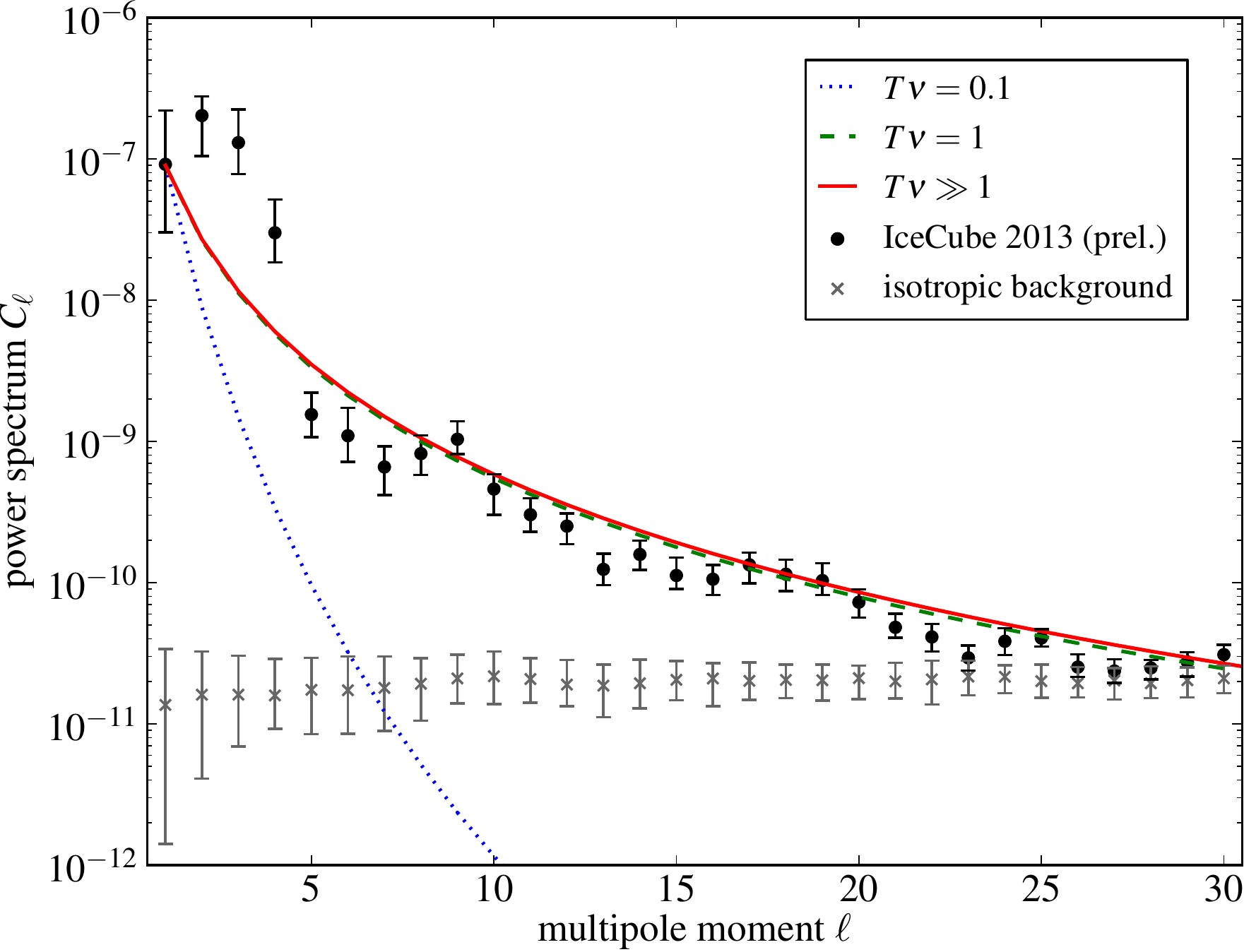}
\caption[]{Angular power spectrum (black dots) at the 68\% confidence level measured with IceCube~\citep{SantanderICRC} at median energy of 20~TeV compared to the model prediction (\ref{eq:Csol}) for $\nu T=0.1$ (blue dotted) and $\nu T=1$ (green dashed) as well as the asymptotic value (\ref{eq:asym}). We also show the power spectrum of scrambled ({\it i.e.}~isotropized) data from Ref.~\citep{SantanderICRC} (gray crosses).}\label{fig1}
\end{figure}

Cosmic ray anisotropies up to the level of one-per-mille have been observed at various energies by the observatories Tibet AS-$\gamma$~\citep{Amenomori:2005dy,Amenomori:2006bx}, Super-Kamiokande~\citep{Guillian:2007}, Milagro~\citep{Abdo:2008kr,Abdo:2008aw}, ARGO-YBJ~\citep{Vernetto:2009xm,ARGO-YBJ:2013gya}, EAS-TOP~\citep{Aglietta:2009mu}, IceCube~\citep{Abbasi:2010mf,Abbasi:2011ai,SantanderICRC} and HAWC~\citep{Abeysekara:2013rka}. The explanation of the strength and phase of the observed dipole anisotropy is challenging, but is qualitatively consistent with the diffusive prediction~\citep{Blasi:2011fm}. However, some of the observations also show significant multi-TeV CR excesses at smaller angular scales with unknown origin. In particular, a high statistics sample of multi-TeV CRs seen by the IceCube observatory~\citep{SantanderICRC} shows significant power in small-scale multipole moments with $\ell\lesssim20$ as shown in Fig.~\ref{fig1}.

It has been speculated that localized CR excesses can be a combined effect of CR acceleration in nearby supernova remnants~\citep{Salvati:2008dx} and the local intergalactic magnetic field structure introducing an energy-dependent magnetic mirror leakage~\citep{Drury:2008ns} or preferred CR transport directions~\citep{Malkov:2010yq}. Magnetic reconnections in the heliotail~\citep{Lazarian:2010sq}, non-isotropic particle transport in the heliosheath~\citep{Desiati:2011xg} or the heliospheric electric field structure~\citep{Drury:2013uka} have also been considered as a source of these small-scale anisotropies. Another variant considers the effect of magnetized outflow from old supernova remnants~\citep{Biermann:2012tc}. More exotic models invoke strangelet production in molecular clouds~\citep{Kotera:2013mpa} or in neutron stars~\citep{Perez-Garcia:2013lza}.

In another recent paper~\citep{Giacinti:2011mz} it was argued that the local turbulent magnetic field configuration within a few scattering lengths from the observer can induce higher multipole moments in the CR arrival direction from the existence of a large scale dipole moment. The authors support this idea via numerical back-tracking of mono-energetic CRs in a particular realization of random fields using a global dipole moment as the initial value. This elegant concept offers the possibility that the study of higher multipole anisotropies can probe the structure of the turbulent magnetic field.

However, a quantitative description of this mechanism has so far not been available. A major challenge consists of an accurate description of the transition region between the diffusive particle transport on large scales and the local deterministic flow of particles where CR back-tracking methods are applied. For the discussion of these second order corrections of the diffusion approximation it is necessary to start from linear Boltzmann equations that describe the evolution of the CR phase space density $f(t,{\bf x},{\bf p})$. In the absence of sources and particle collisions this is simply Liouville's theorem, $\dot{f}=0$. 

We will start in the following with a brief recap of the standard approach of CR diffusion. Using the idealized situation of a homogenous global dipole moment we will then discuss the development of small-scale anisotropies and apply this result to the observed power spectrum of CRs. Throughout this paper we work in Heaviside-Lorentz units and use (hatted) bold-face quantities for (unit)-vectors and tensors.

{\it Diffusion Approximation.}---Before we start the discussion of small-scale anisotropies it is instructive to recall the derivation of the diffusion equation from Liouville's theorem, $\dot{f}=0$. This can be written in the form
\begin{equation}\label{eq:Boltzmann}
\dot{f} = \partial_t f + \dot{\bf x}\nabla_{\bf x} f + \dot {\bf p}\nabla_{\bf p}f  = 0\,.
\end{equation}
In a static magnetic field configuration and neglecting static electric fields the equation of motion of a charged particle with energy $p_0$ is given as $\dot {\bf p} = {\bf p}\times e{\bf B}/p_0$ and $\dot p_0=0$. (For simplicity, we also assume relativistic CRs, $p_0\simeq p$.) As usual, we split the magnetic field into a regular and turbulent component as ${\bf B} = \bar{\bf B} +\mathfrak{B}$. For the following discussion we introduce the angular momentum operator ${\bf L} = -i{\bf p}\times\nabla_{{\bf p}}$ and the rotation vectors ${\boldsymbol \Omega} = e\bar{\bf B}/p_0$ and ${\boldsymbol \omega} = e\mathfrak{B}/p_0$.

Splitting the phase space distribution into $f = \bar f +\mathfrak{f}$ with the magnetic ensemble-average $\bar f = \langle f\rangle$ we can write the Boltzmann equation as~\citep{JonesApJ1990}
\begin{equation}\label{eq:genBoltzmann}
\partial_t\bar f  + \hat{\bf p} \nabla_{\bf x}\bar f  -i\boldsymbol{\Omega}{\bf L}\bar f  = \left\langle i\boldsymbol{\omega}{\bf L}\mathfrak{f}\right\rangle\,.
\end{equation}
The r.h.s.~of Eq.~(\ref{eq:genBoltzmann}) encodes the particle scattering in the random fields. Here and in the following we assume {\it isotropic} turbulence. The influence of the turbulence is typically approximated as a friction term~\citep{Bhatnagar1954} that drives the distribution $\bar{f}$ to its isotropic mean $\phi = \int {\rm d}\hat{\bf p} \bar{f}$ with an effective relaxation rate $\nu$, {\it i.e.}~$\left\langle i\boldsymbol{\omega}{\bf L}\mathfrak{f}\right\rangle \simeq -\nu\left(\bar f-\phi/(4\pi)\right)$. Note that this simple treatment of the scattering term is only approximate. Depending on the strength and coherence length of the turbulent magnetic field, the effective relaxation can in general be anisotropic even for a fully isotropic turbulence. Also, the relaxation time for higher multipole moments is expected to be shorter as we will argue in the following.

For a solution of Eq.~(\ref{eq:genBoltzmann}) one then makes the Ansatz $ f = (\phi + 3\hat{\bf p}{\bf \Phi})/(4\pi)$ where on top of the monopole $\phi$ we assume a non-vanishing dipole component of strength ${\bf \Phi}$. One can show that these functions satisfy
\begin{gather}\label{eq:diff1}
\partial_t\phi  + \nabla_{\bf x}{\bf \Phi} = 0\,, \\
\partial_t{\bf \Phi}+ (1/3)\nabla_{\bf x} \phi +\boldsymbol{\Omega}\times {\bf \Phi}=-\nu{\bf \Phi}\,.\label{eq:diff2}
\end{gather}
In the diffusion approximation it is typically assumed that the dipole component is only slowly varying over the relaxation time $1/\nu$ and hence $\partial_t {\bf \Phi} \simeq 0$. In this case we can rewrite Eq.~(\ref{eq:diff2}) as Fick's law, ${\bf \Phi} = -{\bf K}\nabla_{\bf x}\phi$, and combine Eqs.~(\ref{eq:diff1}) and (\ref{eq:diff2}) into the familiar diffusion equation $\partial_t\phi \simeq \nabla_{\bf x}({\bf K}\nabla_{\bf x}\phi)$ with diffusion tensor
\begin{equation}
{K}_{ij} = \frac{1}{3\nu}\frac{\nu^2\delta_{ij}-\nu{\Omega}_k\epsilon_{ikj}+{\Omega}_i{\Omega}_j
}{\nu^2+\Omega^2}\,.\end{equation}
Its eigenvalues are $1/3\nu$ and $1/3(\nu\pm i\Omega)$, corresponding to diffusion parallel and perpendicular to the magnetic field, respectively. The solution to the diffusion equation with initial condition $\phi(0,{\bf x}) = \delta^{(3)}({\bf x})$ is simply
\begin{equation}\label{eq:diffsol}
\phi(t,{\bf x}) = \frac{1}{(4\pi t)^{3/2}\sqrt{\det{\bf K}}}\exp\left(-\frac{{\bf x}^t{\bf K}^{-1}{\bf x}}{4t}\right)\,.
\end{equation}

{\it Anomalous Anisotropies.}---For the study of small-scale anisotropies it is necessary to start from Liouville's theorem, $\dot{f} = 0$, before ensemble-averaging over the turbulent magnetic field has been applied. The phase-space distribution $f({\bf p}_0)$ at time $t=0$ and ${\bf x} =0$ can then be expressed by the distribution at $t=-T$ as $f({\bf p}_0) = f(-T,{\bf x}(-T),{\bf p}(-T))$ with ${\bf x}(-t) = -\int_0^t{\rm d}t'{\bf p}(-t')$ and ${\bf p}(-t) = {\mathcal{R}}(t){\bf p}|_{{\bf p}={\bf p}_0}$.
Here, we introduced the rotation operator~\cite{Casse:2001be}
\begin{equation}\label{eq:Rop}
{\mathcal{R}}(t) = R(t)\mathcal{T}\left\lbrace \exp\left(i\int\limits_0^t{\rm d}t'\,\boldsymbol{\omega}({\bf x}(-t')){\bf L}(t')\right)\right\rbrace\,,
\end{equation}
where $\mathcal{T}$ denotes time-ordering of the terms in the expansion of the exponential function, $R(t) = \exp(it\boldsymbol{\Omega}{\bf L})$ is the global rotation in the regular magnetic field and ${\bf L}(t') = R^{-1}(t'){\bf L}R(t')$. Generally, the action of the rotation operator (\ref{eq:Rop}) is non-trivial and is typically carried out by numerical studies or in quasi-linear theory, {\it e.g.}~\cite{Casse:2001be}. In the following we will exploit analytic arguments to study the development of small-scale anisotropies. 

Expressing ${\mathcal{R}}(t)$ as a Dyson series and repeatedly expanding $\boldsymbol{\omega}(x)$ as a Fourier series, it is easy to see that the ensemble average of (\ref{eq:Rop}) can be expressed as~\footnote{Here and in the following it is understood that ${\bf L}$ is not acting on ${\bf p}$ in the operator expansion of $\langle{\mathcal{R}}(T)\rangle$.}
\begin{equation}\label{eq:Rdiff}
\langle{\mathcal{R}}(T)\rangle = {R}(T)\,\zeta(T,\lbrace\mathcal{O}_n\rbrace)\,,
\end{equation} 
where $\zeta$ is an operator function that depends on parity-even scalar products $\mathcal{O}_n$ of ${\bf L}$ and $\hat{\bf p}$ and their  projections along ($\parallel$) the regular magnetic field. The perpendicular ($\perp$) contributions can be expressed as ${\bf L}^2_\perp={\bf L}^2-{\bf L}^2_\parallel$, {\it etc.} Motivated by the results of quasi-linear theory we expect that for sufficiently large times the operator function in Eq.~(\ref{eq:Rdiff}) can be approximated by an exponential, $\zeta \simeq \exp(-T(\nu_\perp{\bf L}_\perp^2+\nu_\parallel{\bf L}_\parallel^2)/2)$ with effective relaxation rates $\nu_\parallel$ and $\nu_\perp$. In the case of strong turbulence with $\nu_\perp=\nu_\parallel=\nu$ we recover the diffusion equations (\ref{eq:diff1}) and (\ref{eq:diff2}) for a dipole ${\bf \Phi} = \int {\rm d}\hat{\bf p}\hat{\bf p} {f}$ and ${\bf L}^2=2$. 

Let's assume that at time $t=-T$ the field configuration is spatially homogeneous, $f(-T,{\bf x},{\bf p}) = f(-T,{\bf p})=\sum_{\ell\geq0}\sum_{|m|\leq\ell}\widetilde{a}_{\ell m}Y_{\ell m}(\hat{\bf p})$, with power spectrum $(2\ell+1)\widetilde{C}_\ell = \sum_{|m|\leq\ell}|\widetilde a_{\ell m}|^2$. The configuration at time $t=0$ and ${\bf x}=0$ can then be expressed as $f({\bf p}_0) = {\mathcal{R}}(T)f(-T,{\bf p})|_{{\bf p}={\bf p_0}}$. Note that this configuration is not a stationary solution to the diffusion equations~(\ref{eq:diff1}) and (\ref{eq:diff2}), since it does not obey Fick's law, but it is a physical situation in the sense that we can prepare a system at times $t=-T$ with these properties. We will discuss in the next section how this special situation applies to the more general case of a global stationary solution of the diffusion equation. 

In the following we are interested in the ensemble average of the angular power spectrum. This can be derived via the auto-correlation function defined as
\begin{equation}\label{eq:xi}
\xi(\eta) =  \frac{1}{8\pi^2}\!\int\!{\rm d}\hat{\bf p}_1\!\int\!{\rm d}\hat{\bf p}_2\delta(\hat{\bf p}_1\hat{\bf p}_2-\cos\eta) f({\bf p}_1)f({\bf p}_2)\,.
\end{equation}
One can show that this function can be expressed in terms of Legendre polynomials $P_\ell$ as
\begin{equation}\label{eq:etageneral}
\xi(\eta) = \frac{1}{4\pi}\sum_{\ell\geq0}(2\ell + 1)C_\ell(T) P_\ell(\cos\eta)\,,
\end{equation}
where the $C_\ell(T)$ is the power spectrum of the configuration after it evolved over the time $T$. The ensemble-averaged angular power spectrum can then be inferred from the ensemble-average of the correlation function $\xi(\eta)$. Note that this depends on the average of the product $\langle f({\bf p}_1)f({\bf p}_2)\rangle$, instead of the product of averages $\langle f({\bf p}_1)\rangle\langle f({\bf p}_2)\rangle$. {\it This is the source for anomalous anisotropies.} 

The necessary development of small-scale anisotropies can be most easily derived from the behavior of the correlation function (\ref{eq:xi}) in the case $\eta=0$, {\it i.e.}~for identical particle flow, ${\bf p}_1(-t) = {\bf p}_2(-t)$. In this case the correlation function reduces to 
\begin{equation}
\xi(0) = \frac{1}{4\pi}\int{\rm d}\hat{\bf p}_1f^2(-T,{\bf p}_1(-T))\,.
\end{equation}
This expression is for a particular magnetic field ensemble. However, for the ensemble average we can make use of the fact that the combination of a regular and average random rotation of an isotropic distribution is eventually isotropic. More precisely, in the case of pure turbulence the average distribution is isotropic at all times, whereas for the combination of turbulent and regular magnetic fields the average distribution at intermediate times can in principle be non-isotropic but is expected to reach an isotropic state eventually. Using the orthogonality of spherical harmonics the average is hence
\begin{equation}\label{eq:eta0}
\langle\xi\rangle(0) = \frac{1}{4\pi}\sum_{\ell\geq0} (2\ell+1)\widetilde{C}_\ell\,.
\end{equation}
Comparing Eq.~(\ref{eq:etageneral}) with Eq.~(\ref{eq:eta0}) and using $P_\ell(1)=1$ we can derive the identity
\begin{equation}\label{eq:identity}
\sum_{\ell\geq0}(2\ell+1)\widetilde{C}_\ell = \sum_{\ell\geq0}(2\ell+1)\left\langle C_\ell(T)\right\rangle\,.
\end{equation}
{\it This implies that for initial conditions $\widetilde{C}_1 > 0$ and $\widetilde{C}_\ell =0 $ for $\ell>1$ the power spectrum at later times $T$ will develop higher multipole moments if $\left\langle C_0(T)\right\rangle = \widetilde{C}_0$ and $\left\langle C_1(T)\right\rangle < \widetilde{C}_1$.}

For general correlation angles $\eta$ the ensemble-averaged correlation function (\ref{eq:xi}) can be written as
\begin{widetext}
\vspace{-1em}
\begin{equation}\label{eq:master}
\langle\xi\rangle(\eta) = \frac{1}{4\pi}\sum_{\ell\geq0} P_\ell(\cos\eta)\!\int\!{\rm d}\hat{\bf p}_1\!\int\!{\rm d}\hat{\bf p}_2\!\sum_{m=-\ell}^\ell\!{Y}^*_{\ell m}(\hat{\bf p}_1)Y_{\ell m}(\hat{\bf p}_2)\langle{\mathcal{R}}_1(T){\mathcal{R}}_2(T)\rangle f(-T,{\bf p}_1)f(-T,{\bf p}_2)\,,
\end{equation}
\end{widetext}
where the rotation operator ${\mathcal{R}}_{i}(T)$ is only acting on ${\bf p}_i$.
Similar to Eq.~(\ref{eq:Rdiff}) we can express the ensemble-averaged operator product in Eq.~(\ref{eq:master}) as 
\begin{equation}\label{eq:opfunc}
\langle{\mathcal{R}}_1(T){\mathcal{R}}_2(T)\rangle = {R}_1(T){R}_2(T)\,\chi(T,\lbrace\mathcal{O}_n\rbrace)\,,
\end{equation}
where ${R}_i$ ($i=1,2$) are again global rotation operators and $\chi$ can be expanded as a series of operators $\mathcal{O}_n$ that are scalar products of $\hat{\bf p}_i$ and ${\bf L}_i$ as well as their projection along the magnetic field. The terms $\mathcal{O}_n$ are parity even (${\bf p}_i\leftrightarrow-{\bf p}_i$) and symmetric (${\bf p}_1\leftrightarrow{\bf p}_2$). In the degenerate case $\eta=0$ with ${\bf p}_1 = {\bf p}_2 = {\bf p}$ the function $\chi$ can only depend on operator products of the sum ${\bf J}={\bf L}_1+{\bf L}_2$ and hence the only possible operators $\mathcal{O}_n$ are those combinations appearing in Eq.~(\ref{eq:Rdiff}). Note that for a negligible regular magnetic field this list reduces to ${\bf J}^2$ and for ${\bf J}^2=0$ we have simply $\chi=1$. We will show in the following that we recover Eq.~(\ref{eq:eta0}) from this case. For $\eta>0$ we expect that the rotations of ${\bf p}_1$ and ${\bf p}_2$ are uncorrelated at sufficiently large times and hence 
\begin{equation}\label{eq:decoh}
\lim_{T\to\infty}\langle{\mathcal{R}}_1(T){\mathcal{R}}_2(T)\rangle = \langle{\mathcal{R}}_1(T)\rangle\langle{\mathcal{R}}_2(T)\rangle\,.
\end{equation}
However, the smaller $\eta$ the longer it takes that this asymptotic regime is reached and this $\eta$-dependence introduces higher multipole moments. 

For simplicity, we assume in the following that the turbulence is strong such that the effect of the regular magnetic field can be ignored. For fixed $\ell_1$ and $\ell_2$ we can express the sum over $m_1$ and $m_2$ in the multipole expansion of $f(-T,{\bf p}_1)f(-T,{\bf p}_2)$ as eigenstates of $J$ and $M$ of total angular momentum ${\bf J}={\bf L}_1+{\bf L}_2$ using Clebsch-Gordan coefficients. Now, the integral over the sum $\Psi_\ell=\sum_{|m|\leq\ell} Y^*_{\ell m}(\hat{\bf p}_1)Y_{\ell m}(\hat{\bf p}_2)$ in Eq.~(\ref{eq:master}) projects onto $J=0$ eigenstates with $M=0$. Hence, we recover Eq.~(\ref{eq:eta0}) in the case $\eta=0$. For general $\eta$ and in the case of strong turbulence we can express the asymptotic value of $\chi$ as a function of ${\bf L}_1^2$ and ${\bf L}_2^2$ and the asymptotic ensemble average takes on the particularly simple form of
\begin{equation}\label{eq:avxi_special}
\lim_{T\to\infty}\langle\xi\rangle(\eta) \simeq \frac{1}{4\pi}\sum_{\ell\geq0}\chi_\ell(\eta,T)(2\ell+1)\widetilde{C}_\ell P_\ell(\cos\eta) \,,
\end{equation}
with $\chi_{0}=1$ and $\ell_1=\ell_2=\ell$. From Eq.~(\ref{eq:decoh}) we expect that for large times $T$ the multipole anomalies behave as
\begin{equation}
\lim_{T\to\infty}\chi_\ell(\eta,T) \propto \exp(-T \ell(\ell+1)\nu)\,,
\end{equation}
where the extra factor $\ell(\ell+1)$ accounts for the ${\bf L}^2$-dependence of higher multipole correlation functions. In particular, $\langle C_0\rangle\simeq\widetilde{C}_0$ as well as $\langle\dot{C}_1\rangle < 0 $ and hence Eq.~(\ref{eq:identity}) requires that the power spectrum generates higher multipole moments.

{\it Cosmic Ray Anisotropies.}---In the previous section we have assumed a specific situation of a homogenous initial condition, $f(-T,{\bf x},{\bf p}) = f(-T,{\bf p})$. This does not correspond to a stationary diffusion solution as one can see from the inspection of Eq.~(\ref{eq:diff2}). However, since the average diffusive propagation distance according to Eq.~(\ref{eq:diffsol}) is $\langle x\rangle \sim \sqrt{T/\nu}$ we see that the spatial gradient term ${|\bf x}\nabla_{\bf x}f|\sim 3\nu\langle x\rangle|{\bf \Phi}|/(4\pi)$ is small compared to the dipole gradient $|{\bf p}\nabla_{\bf p} f| \sim 3|{\bf \Phi}|/(4\pi)$ as long as $T\nu\lesssim1$. As we saw in the previous section, the relaxation rate of higher multipole moments is $\propto\ell(\ell+1)\nu$ and if these are induced by a global diffuse dipole moment, they have to be generated {\it locally} within times $T\nu\lesssim1$.

From the previous discussion we expect that the strength of individual multipoles eventually decreases in time, $\langle \dot{C}_\ell\rangle<0$ for $\ell\geq1$, and that the individual multipoles follow eventually $\langle\dot{C}_\ell\rangle \simeq -\nu_\ell\langle{C}_\ell\rangle$ with $\nu_\ell\simeq \ell(\ell+1)\nu$. On the other hand, Eq.~(\ref{eq:identity}) implies that $\sum_{\ell\geq0}(2\ell+1)\langle\dot{C}_\ell\rangle=0$ in the case of strong turbulence. Assuming a {\it linear} multipole evolution this implies a coupled set of differential equations of the form
\begin{equation}\label{eq:model}
\langle\dot{C}_{\ell}\rangle \simeq -\nu_\ell\langle{C}_\ell\rangle + \sum_{\ell'\geq0}\nu_{\ell'\to\ell}\frac{2\ell'+1}{2\ell+1}\langle{C}_{\ell'}\rangle\,,
\end{equation}
where $\nu_\ell = \sum_{\ell'\geq0} \nu_{\ell\to\ell'}$ and $\nu_{\ell\to\ell'}$ is the transition rate between multipoles $\ell'$ and $\ell$.

To proceed further, we assume that {\it (i)} the multipole generation is hierarchical, {\it i.e.}~$\nu_{\ell'\to\ell} \simeq 0$ for $\ell'>\ell$, and {\it (ii)} the generation is dominated by transitions between consecutive multipoles, {\it i.e.}~$\nu_{\ell}\simeq \nu_{\ell\to\ell+1}$. Under these conditions the solution to Eq.~(\ref{eq:model}) for initial condition $\langle\dot{C}_{\ell}\rangle=0$ for $\ell\geq2$ and $\langle\dot{C}_{1}\rangle=\widetilde{C}_1$ is simply~\citep{Ahlers:2010ty}
\begin{equation}\label{eq:Csol}
\langle{C}_{\ell}\rangle(T) \simeq \frac{3\widetilde{C}_1}{2\ell+1}\prod_{m=1}^{\ell-1}\nu_m\sum_n \prod_{p=1(\neq n)}^\ell\frac{e^{-T\nu_n}}{\nu_p-\nu_n}\,.
\end{equation}
Remarkably, from these simple assumptions we can derive a stationary solution for the multipole ratios,
\begin{equation}\label{eq:asym}
\lim_{T\to\infty}\frac{\langle{C}_{\ell}\rangle(T)}{\langle{C}_{1}\rangle(T)} \simeq \frac{18}{(2\ell+1)(\ell+2)(\ell+1)}\,.
\end{equation}
The solution (\ref{eq:Csol}) and its asymptotic ratio (\ref{eq:asym}) are shown in Fig.~(\ref{fig1}) in comparison with the power spectrum observed with IceCube~\citep{SantanderICRC}. The scaling of $\ell>5$ multipoles is well described by the asymptotic form (\ref{eq:asym}). 

There is significantly more power in the lower $2\leq\ell\leq4$ multipoles. Generally, the systematic uncertainty of the power spectrum (not shown in Fig.~\ref{fig1}) is expected to be larger for low-$\ell$ multipoles due to the partial sky coverage $f_{\rm sky}\simeq 1/3$ of IceCube's power spectrum analysis~\citep{SantanderICRC} in combination with cosmic variance. The relative uncertainty of the induced power spectrum ($\ell\geq2$) can be estimated as $\langle(\Delta C_\ell)^2\rangle/\langle C_\ell\rangle^2 \sim 2/(2\ell+1)/f_{\rm sky}$ assuming a Gaussian distribution of the individual $a_{\ell m}$'s~\cite{Chon:2003gx}. The systematic error of the dipole is expected to be dominated by its coupling to the quadrupole due to the partial sky coverage. Additional contributions to the low-$\ell$ multipoles can be expected from non-turbulent contributions to a large scale anisotropy, for instance the effect of the heliosphere described in the introduction. At large values $\ell\gg20$ the observed multipoles become indistinguishable from an isotropic distribution.

{\it Summary.}---We have discussed the generation of small-scale anisotropy from the local turbulence of magnetic fields. Based on Liouville's theorem we could show that for an idealized situation of a homogenous large scale anisotropy the total sum of multipoles is conserved, which implies the generation of small-scale anisotropy. We have applied this result to the power spectrum observed in multi-TeV CRs and could show that the relative decrease of the power spectrum of medium scale anisotropies is well described by this model.

A specific prediction of this model is that the relative power spectrum of medium scale anisotropies is asymptotically independent of the diffusive relaxation rate $\nu$ and is hence not expected to show a strong dependence on the type of diffusion. Only the absolute scale of the dipole anisotropy reveals the dependence on CR rigidity or magnetic turbulence. Note, however, that any actual anisotropy measurement averages over CR energies. The angular correlation of CRs with different rigidities is expected to damp the $\langle C_\ell\rangle$ spectrum, possibly also depending on angular scale and magnetic turbulence. Finally, the assumptions leading to Eq.~(\ref{eq:Csol}) will not necessarily capture all possible effects of magnetic turbulence. These questions will be further studied in a forthcoming paper~\citep{AhlersSantander}.
 
{\it Acknowledgements.}---The author would like to thank Marcos Santander for discussions and providing the data points of Fig.~\ref{fig1}. This work is supported by the National Science Foundation under grants OPP-0236449 and PHY-0236449.

\end{document}